\documentclass[aps,showpacs,preprintnumbers,amsmath,amssymb, 
eqsecnum,
twocolumn, tightenlines
]{revtex4}

\usepackage[dvips]{graphicx}
\usepackage{bm}

\begin{document}

\bibliographystyle{apsrev} 

\title {Fermions in energon field}

\author{M. Yu. Kuchiev} \email[Email:]{kmy@phys.unsw.edu.au}

\affiliation{School of Physics, University of New South Wales, Sydney
  2052, Australia}

    \date{\today}
    %\date{25 October, 2014}

\begin{abstract}
%% Text of abstract
The behavior of fermions in the gauge field created by the energon, 
a recently found classical solution of the non-Abelian gauge theory,
is considered.
The spectrum of fermions is evaluated explicitly for the case when 
parameters governing the energon are chosen to make it look
similar to the chain of separated instantons and antiinstantons. 
The zero eigenvalue in the spectrum of the Dirac operator is found. 
The role of a specific discrete symmetry, which distinguishes the energon, is discussed.
\end{abstract}

    \pacs{11.15.-q, %Gauge field theories
          11.15.Kc %Classical and semiclassical techniques
          }  
    \maketitle

\section{Introduction}
\label{Introduction}

The semiclassical approach to quantum Yang-Mills field is known to describe a number of its important properties.
This approach is based on classical solutions. 
An important class of these solutions is the self-dual solutions. 
The firstly discovered  BPST instanton of Ref.\cite{Belavin:1975fg} remains one of the most useful tools, 
for a review see e.g. Ref.\cite{Prasad:1980yy}.
The general multi-instanton solution is described by the ADHM construction of Ref.\cite{Atiyah:1978ri}.

A modification of the instanton solution, which is known as the caloron of Refs. \cite{Harrington:1978ve,Harrington:1978ua}, 
has a periodic nature that makes it interesting for applications at finite temperatures.
Recent developments and further references for the caloron can be found in reviews Refs. \cite{vanBaal:2009pn,Weinberg:2006rq}.
The periodic conditions were differently implemented in the so called periodic instantons of Refs. \cite{Khlebnikov:1991th}; the configuration proved useful in multiparticle production and studies of the barion and lepton number violation. %\cite{Bezrukov:2003qm,Bezrukov:2003er,Rubakov:1996vz}.

Ref.\cite{Klinkhamer:1984} introduced the sphaleron, a self-consistent static solution 
for interacting
gauge and Higgs fields. It was argued in Ref. \cite{Ostrovsky:2002cg} that the pure
gauge field can exhibit the sphaleron-type behavior when it
is treated using particular restrictions. The found gauge
field configuration, called COS-sphaleron, possesses only the magnetic field.   

Another line for generalization of classical solutions was pursued in Refs. \cite{Kuchiev:2009rz,Kuchiev:2012epl}.
The idea was to take a minimum of the classical action under the condition that 
the Euclidean energy is fixed; the corresponding field configuration was called the energon. 
Hence the energy plays the role of an additional parameter, 
which can be tuned to adjust properties of the energon to the needs of a problem at hand.
In \cite{Kuchiev:2009rz} it was shown that for finite temperatures the energon provides 
exponential enhancement for the probability of tunneling through the potential barrier 
separating regions with different topological charges.

A semiclassical solution in gauge theory usually incorporates strong gauge fields.
It is important therefore to consider their influence on fermions, 
which may have interesting consequences.
For example, for instantons
the fermionic zero modes
have valuable physical manifestations
\cite{tHooft1976,tHooft1976PRL};
they also give an interesting example of the application of
the Atiyah-Singer theorem \cite{Atiyah-Singer-1963}.

%The instantons give a good example. 
%Alongside physical valuable manifestations
%\cite{tHooft1976,tHooft1976PRL} their
%fermionic zero modes provide 
%an interesting example of the application for
%the Atiyah-Singer theorem \cite{Atiyah-Singer-1963}.

The present work considers the behavior of fermions in the field of the energon.
In order to simplify the problem the parameters governing the 
energon configuration are tuned in such a way that
the gauge field of the energon looks similar to the 
field created by a periodic chain of well separated instantons 
and antiinstantons. The small parameter $\beta=\rho/T\ll 1$, where $\rho$
is the radius of the mentioned instantons and antiinstantons and $T$ is their
separation, is used to solve the problem analytically.
A continuum spectrum of eigenstates of the Dirac operator is
found explicitly. 

The corresponding eigenvalues of the Dirac operator are small,
which agrees with the fact that the field of the energon chosen 
resembles a field of well separated instantons and antiinstantons.
However, generally these eigenvalues differ from zero. 
Interestingly, there exist two special states,
which are annihilated  by the Dirac operator. In other words,
the eigenvalue of the Dirac operator for these two states turns 
zero, and hence the energon possesses the fermionic zero modes.
%This fact is verified here for the case $\beta\ll 1$.
%It remains to be seen whether its validity can be justified in general case. 

\section{Fermions in energon field}
\label{Massless fermions}
We presume that the gauge group is SU(2), fermions belong
to its fundamental representation; Euclidean formulation of the theory
is chosen. Consider the Dirac equation 
for massless fermions
\begin{equation}
i\, \gamma_\mu\nabla_\mu\,\psi\,=\,\varepsilon\,\psi~.
\label{dirac}
\end{equation}
Here  $\varepsilon$ and $\psi$ play the roles of 
the eigenvalue and corresponding eigenfunction of the Dirac operator,
which in the Euclidean formulation is Hermitian. In this Section 
we will formulate the eigenproblem for Eq.(\ref{dirac}) 
when the potential $A^a_\mu$, 
which appears in 
the covariant derivative $\nabla_\mu=\partial_\mu-i A^a_\mu \tau^a/2 $, 
is created by the energon.

To clarify notation note that we use the Euclidean Dirac matrices $\gamma_\mu$ 
defined as follows
\begin{equation}
\gamma_\mu\,=\,
\left(
\begin{array}
{cc}
0 & \sigma_\mu^+
\\
\sigma_\mu^- & 0
\end{array}
\right)\,,
\label{gamma}
\end{equation}
where $\mu=1,...\,4$ and the Pauli matrices read
\begin{equation}
\sigma_\mu^{\pm}\,=\,(\pm i\, \boldsymbol{\sigma},1)~.
\label{sigma}
\end{equation}
We need first
to outline basic properties of
the energon solution of \cite{Kuchiev:2009rz,Kuchiev:2012epl}.  The energon provides a conditional 
minimum for the Euclidean action achieved when
the Euclidean energy is fixed. 
The classical action $S$ and Euclidean energy $\mathcal{E}$ 
of the SU(2) gauge field read
\begin{align}
S  \,=\,\int \, (K+V)\,d\tau~,\quad
%\label{action}
\mathcal{E}\,=\,K-V~.
\label{action-energy}
\end{align}
Here
\begin{align}
K=\frac{1}{2g^2} \int
\mathbf{E}^a\cdot \mathbf{E}^a d^3r, \quad
V=\frac{1}{2g^2} 
\int \mathbf{B}^a\cdot \mathbf{B}^a d^3r
\label{KU}
\end{align}
are expressed via the electric $E^a_m=F^a_{m4}$ and magnetic $B^a_m=\frac 1 2 \epsilon_{mnl} F^a_{nl}$ components of the gauge field $F_{\mu\nu}^a$ ($a=1,2,3$ is an isotopic index;
$m,n,l=1,2,3$ are indexes of the 3D coordinate space, whose
presence is replaced in (\ref{KU}) by bold and dot-product notation).
The signs chosen in front of $V$ in Eqs.(\ref{action-energy}) comply with the
Euclidean description.

As was mentioned, the energon provides the minimum for the action and hence satisfies
\begin{equation}
\delta S\,=\,0~.
\label{minS}
\end{equation}
The minimum is considered under the restriction that $\mathcal{E}$ takes a chosen, fixed value. Hence
$\mathcal{E}$ can be considered as an additional parameter.
%(which can be used to adjust properties of the energon to a particular problem at hand).

Equations (\ref{action-energy})-(\ref{minS}) have a surprisingly simple solution 
\cite{Kuchiev:2009rz,Kuchiev:2012epl}. One considers 
firstly a non-restricted solution of (\ref{minS}), which can be chosen as any self-dual, or anti-selfdual solution $A^a_{\mu,\mathrm{SD}}(x)$. Secondly, alongside
conventional Euclidean coordinates $x_\mu=({\bf r},\tau)$ one introduces also
the modified coordinates
\begin{equation}
z_\mu=({\bf r}, q(\tau))~,
\label{x to z}
\end{equation}
in which the Euclidean time $\tau$ is replaced by a function $q(\tau)$ 
specified below, see Eq.(\ref{q dot squared}). 

We can write down 
the energon solution of Eqs. (\ref{action-energy})-(\ref{minS})
as follows
\begin{equation}
A_m^a(x)\,=\,A_{m,\mathrm{SD}}^a(z)~,
\quad~
A_4^a(x)\,=\,A_{4,  \mathrm{SD} }^a(z)\,\dot{q}(\tau)~,
\label{A-energon}
\end{equation}
where $\dot{q}(\tau)=\frac{dq}{d \tau}(\tau)$. Note the difference in coordinates in the left and right hand sides here.
The function $q(\tau)$ should be found from the condition
specifying the energy, which reads
\begin{equation}
\mathcal{E}=
(\dot q^2-1)\frac 1{2g^2}
\int \mathbf{B}^a_{\mathrm{SD}}(z)
\cdot \mathbf{B}^a_{\mathrm{SD}}(z) d^3r\,.
\label{q dot squared}
\end{equation}
The integral in the right-hand side here is a function of $q$, which
is defined by the chosen self-dual solution. Consequently, 
Eq.(\ref{q dot squared}) can be considered an ordinary differential equation 
on $q(\tau)$. Its solution finalizes description of the energon
potential in Eq.(\ref{A-energon}).
%, see \cite{Kuchiev:2009rz,Kuchiev:2012epl} for further details.

Thus for each self-dual solution $A_{\mu,\mathrm{SD}}^a$
there exists a set of energon
solutions $A_{\mu}^a$ parametrized by a value of $\mathcal{E}$. 
In particular, for $\mathcal{E}=0$ the energon is reduced to the 
underlying self-dual solution,
as can be seen from
Eq.(\ref{q dot squared}), which in this case gives
$q(\tau)=\tau$; consequently one deduces
that $x_\mu=z_\mu$, and $A_\mu(x)=A_{\mu,\mathrm{SD} }(x)$.

We will restrict further discussion to the case of
negative $\mathcal{E}$.
Equation (\ref{q dot squared}) ensures in this case 
that $\dot q^2<1$. 
Assuming that the self-dual solution is well 
localized in 4D Euclidean space one observes also that the integral
in Eq.(\ref{q dot squared}) decreases at large $q$.
These two conditions together indicate that
$q(\tau)$ is an oscillating function of $\tau$.
The variation of $\tau$ over the
period $T$
results in the variation of $q(\tau)$ between
its minimum and maximum,
where $\dot q=0$.
An explicit expression for the period can be written in the form
\begin{equation}
T\,=\,\oint \frac{dq}{\dot q}~,
\label{period-T}
\end{equation}
where $\dot q$ should be taken from (\ref{q dot squared}).

Thus, under conditions specified above the energon potential
in Eq.(\ref{A-energon}) and
corresponding field $F^a_{\mu\nu}$ are 
periodic  functions of $\tau$ with the period $T$.
At the same time
$F^a_{\mu\nu}$ is well localized as a function of 3D coordinates ${\bf r}$,
$F^a_{\mu\nu}\rightarrow 0$ when ${\bf r} \rightarrow \infty$.

Returning to the eigenvalue problem in Eq.(\ref{dirac})
we can now be more specific about properties of $\psi$.
Firstly, the localization of the energon in $\bf r$
makes it possible, and obviously appealing, to restrict
our consideration to those $\psi=\psi({\bf r},\tau)$, which are
localized functions of $\bf r$,
\begin{equation}
\psi({\bf r},\tau)  \rightarrow 0\,  ,  \quad {\bf r} \rightarrow \infty~.
\label{r infty}
\end{equation}
Secondly, we need to take into account that being periodic in $\tau$, the energon potential
makes the Dirac operator in Eq.(\ref{dirac}) a periodic operator as well.
Consequently, we can impose on $\psi$ the following condition
\begin{equation}
\psi({\bf r},\tau+T)\,=\,e^{i\omega T} \psi({\bf r},\tau)\,.
\label{Bloch}
\end{equation}
It represents for the case at hand Bloch's theorem, which is commonly used 
in description of crystals in condense matter, see e.g. Ref. \cite{Ashcroft-Mermin}.
Equation (\ref{Bloch}) makes it clear that the parameter 
$\omega$, which specifies properties of the solution,
can be chosen to satisfy the following condition
\begin{equation}
-1\,\le\,\omega\,T/\pi\le\, 1~.
\label{Brillouin}
\end{equation}
This region of $\omega\,T$ can be called the first Brillouin zone having in mind 
similar conditions specifying the zone structure in crystals \cite{Ashcroft-Mermin}.

Summarizing, Eqs. (\ref{dirac}), (\ref{r infty}) and  (\ref{Bloch})
formulate the eigenvalue problem for massless fermions in the energon field
defined by Eqs. (\ref{x to z}), (\ref{A-energon}) and  (\ref{q dot squared}).
According to Bloch's theorem (\ref{Bloch}) the eigenfunction and eigenvalue 
depend on the parameter $\omega$, which belongs to the first Brillouin 
zone (\ref{Brillouin}). The $\omega$-dependence will be marked by a subscript, 
$\psi=\psi_\omega$ and  $\varepsilon=\varepsilon_\omega$.

\section{Discrete R-symmetry}
Consider an important  discrete symmetry that exists in the problem.
Its presence can be described in general terms.
However, in order to simplify presentation 
we restrict our discussion to 
the energon, which is constructed from the BPST instanton. The instanton
potential reads
\begin{equation}
A^a_{\mu, \mathrm{in}}(x)\,=\,\bar \eta^a_{\mu\nu} \frac{x_\nu \tau^a}{x^2(x^2+1)}~.
\label{instanton}
\end{equation}
Here $\bar \eta^a_{\mu\nu}$ are the t'Hooft symbols, t'Hooft singular gauge is presumed,
the instanton center is chosen at the origin, and units are scaled to make the instanton radius unity.
It is useful to present the instanton potential from (\ref{instanton}) in a more detailed 3D notation
\begin{equation}
{\bf A}_{\mathrm{in}}({\rm r},\tau)\!=\!\frac{{\bf r}\!\times\! \boldsymbol{\tau} -\tau\, \boldsymbol{\tau} }{x^2(x^2+1)},
\quad
A_{ 4,\mathrm{in} }( {\rm r} ,\tau)\!=\!\frac{  {\bf r} \cdot \boldsymbol{\tau} }{x^2(x^2+1)}.
\label{3D-ins}
\end{equation}
To avoid confusion, $\tau$ is the Euclidean time and $x^2=r^2+\tau^2$, whereas the bold $\boldsymbol{\tau} $ represents a triplet of the Pauli matrices.
The antiinstanton solution can also be described by (\ref{instanton}), provided $\bar \eta^a_{\mu\nu}$ 
is replaced there by $\eta^a_{\mu\nu}$. 
The known consequence of this fact is that the antiinstanton potential can be obtained from the instanton one by 
applying either the operator $\hat P$ of the inversion of coordinates ${\bf r}$, or the operator $\hat T$ of the reversal of time $\tau$.
As a result the instanton potential remains invariant under a combined influence of $\hat P\,\hat T$,
\begin{equation}
\hat P\,\hat T\,A^a_{\mu, \mathrm{in}}\,=\,A^a_{\mu, \mathrm{in}}~.
\label{PT}
\end{equation}
Let us show that there exists a similar discrete symmetry for the energon solution.
Following the procedure outlined in Section \ref{Massless fermions}
we can build the energon potential starting from the instanton solution (\ref{3D-ins}) as follows
\begin{equation}
{\bf A}({\rm r},\tau)\!=\!\frac{{\bf r}\times \boldsymbol{\tau} -q(\tau)\, \boldsymbol{\tau} }{z^2(z^2+1)},
\quad
A_{ 4}( {\rm r} ,\tau)\!=\!\frac{  \dot q(\tau) \, ({\bf r} \cdot \boldsymbol{\tau}) }{z^2(z^2+1)}.
\label{3D-energon}
\end{equation}
Here $z^2=r^2+q^2(\tau)$.
To fully describe the energon one needs to resolve  Eq.(\ref{q dot squared}) on $q(\tau)$. An additive constant
arising during this procedure can be chosen in such a way as to satisfy $q(0)=0$. After that it is easy to see
that the periodic function $q(\tau)$ changes the sign over the half-period
\begin{equation}
q(\tau+T/2)\,=\,-q(\tau)~,\quad \dot q(\tau+T/2)\,=\,-\dot q(\tau)~.
\label{q=-q}
\end{equation}
Consider now the following operator
\begin{equation}
\hat R\,=\,\hat P \, e^{i \,(T/2)\,\partial_\tau}~.
\label{R-operator}
\end{equation}
Here 
%$\hat P$ is the operator of inversion, which
%for any vector function ${\bf f}({\bf r})$ gives $\hat P \,{\bf f}({\bf r})=-{\bf f}(-{\bf r})$, while 
$\exp({i \,(T/2)\,\partial_\tau})$ is the operator of a shift of the argument  $\tau$ of any function $\phi(\tau)$ 
by the half-period of the energon, $\exp({i \,(T/2)\,\partial_\tau})\,\phi(\tau)=\phi(\tau+T/2)$. 
%This operator satisfies
%\begin{equation}
%\hat R^2\,=\,e^{i \,T\partial_\tau}\,\equiv\,1~,
%\label{R-operator}
%\end{equation}
%where the last identity presumes that $\hat R^2$ is applied to the 
%periodic in  $\tau$ functions, $\phi(\tau+T)=\phi(\tau)$.
%It follows from (\ref{R-operator}) that the eigenvalues $R$ of the operator 
%$\hat R$ equal $R=\pm 1$.
Remembering Eqs.(\ref{q=-q}) one finds that the energon potential defined in (\ref{3D-energon})
is $R$-invariant
\begin{equation}
\hat R\,A_\mu^a(x)\,=\,A_\mu^a(x)~.
\label{R-invariant}
\end{equation}
Observe a similarity with Eq.(\ref{PT}). The difference is that 
instead of the time inversion, which appears in (\ref{PT}) for the instanton case, 
Eq.(\ref{R-invariant}) incorporates the shift of $\tau$ by $T/2$.

The invariance of the energon potential under $\hat R$ 
makes the Dirac operator invariant as well
\begin{equation}
\hat R \,i \,\gamma_\mu\,\nabla_\mu\,\psi\,=\,i\,\gamma_\mu\,\nabla_\mu \hat R \,\psi~.
\label{nabla-invariant}
\end{equation}
As a result the eigenstate $\psi$ of the Dirac operator should also
be an eigenstate of $\hat R$, $\hat R\,\psi=\lambda\, \psi$, where $\lambda$ is
the eigenvalue. The definition of $\hat R$ (\ref{R-operator}) implies
$\hat R^2\,=\,\exp(i\,T\,\partial_\tau)$, where the resulting operator
describes a shift of $\tau$ by the  energon period $T$. 
Remember that this operator is also present in Bloch's periodic conditions
(\ref{Bloch}) which can be written as 
$\exp(i\,T\,\partial_\tau)\,\psi(\tau)\,=\,\exp(i\,\omega\,T)\,\psi(\tau+T)$.
We derive from this fact that $\lambda$ can take only the following values, 
$\lambda\,=\,\pm\,\exp(i\,\omega\,T/2)$. 

Thus, the function $\psi$ can be characterized
by two parameters, one is $\omega$, another is the sign, which 
distinguishes two different values of $\lambda$ for a given $\omega$.
It makes sense therefore to mark the fermion function by these parameters,
$\psi=\psi_{\omega,\pm}$ presuming that its transformation under $\hat R$
reads
\begin{equation}
\hat R\,\psi_{\omega,\pm}\,=\,\pm\,\exp(i\,\omega\,T/2)\,\psi_{\omega,\pm}~.
\label{omega,pm}
\end{equation}
Observe that Bloch's periodic condition (\ref{Bloch}) 
can be derived from (\ref{omega,pm}) because, as was mentioned,
$\hat R^2=\exp(i\,T\,\partial_\tau)$.
We will see below that (\ref{omega,pm})  has a strong impact 
on the fermion spectrum.

\section{Low Euclidean energies}

Consider the case where Euclidean energy is negative and small
$\mathcal {E}<0$, $a |\mathcal {E}|\ll 1$, where $a$ is a typical
distance at which the gauge field varies substantially.
Equation (\ref{period-T}) shows that the period $T$ in this case is large.
To be specific we take the energon, whose vector potential 
is written in (\ref{3D-energon}).
It was found in \cite{Kuchiev:2009rz}
that the relation between the energon period $T$ and small Euclidean energy $\mathcal {E}$ in this case
reads
\begin{equation}
\frac{T}{2\rho} \,\approx\, 2.51 \Bigg(-\frac{3\pi^2 }{g^2 \rho \,\mathcal{E} }  \Bigg)^{1/5}~.
%\,\propto\,(-\mathcal{E})^{-1/5}~.
\label{T/2}
\end{equation}
Here $g$ is the coupling constant of the SU(2) gauge theory. 
Previously we chose the instanton radius as unity, $\rho=1$, 
but here present it explicitly to make formulas more transparent. 
For small energies, $\rho \,|\mathcal{E}|\ll 1$,  Eq.(\ref{T/2}) implies 
\begin{equation}
\beta\,=\,{\rho}/{T}\,\ll\,1~.
\label{beta}
\end{equation}
%This small parameter is used below to simplify analytical calculations.
Observe that for small $\beta$ the gauge field produced by the energon can be strong only
if $\tau$ satisfies either condition $|\tau-s\,T|\lesssim \rho$,  or $|\tau-(s+\frac12) T|\lesssim \rho$ where
$s=0,\pm 1\dots$. In the first case, $|\tau-s\,T|\lesssim \rho$, the field created by the
energon is close to the field of the instanton located at $x_\mu=(0,0,0,s\,T)$.
In the second case, $|\tau-(s+\frac12)\,T|\lesssim \rho$, the energon field is close to the 
field created by the antiinstanton located at $x_\mu=(0,0,0,(s+1/2)\,T)$.

Correspondingly, at small negative $\mathcal {E}$ the energon field can be approximated
by the field of a chain of alternating instantons and antiinstanton, which is stretched along 
the $\tau$-axis being also periodic in $\tau$.
The separation between the closest pair of instantons or antiinstantons equals the energon period $T$,
while the nearest instanton and antiinstanton are separated by $T/2$.
The radius, as well as the orientation of all instantons and antiinstantons
in this chain are the same.

Remember that the pure instanton and pure antiinstanton configurations possess fermionic zero modes. 
This implies that any field configuration
that includes well separated instantons and antiinstantons should
possess fermion modes, which satisfy the Dirac equation (\ref{dirac}) with small
eigenvalues $\varepsilon$, $|\varepsilon| \rho\ll 1$.

Let us find these eigenvalues and the
corresponding eigenfunctions $\psi$ for the energon (\ref{3D-energon}), assuming that
Eq.(\ref{beta}) is valid.
We can apply the method of ``strong coupling'',
which is similar to the one  used conventionally
in solid state physics, see e.g. Section 10 in \cite{Ashcroft-Mermin}.
It is commonly applied for constructing the wave functions of electrons propagating 
in crystals from the wave functions of  atomic electrons, which
are localized on individual atoms. The smaller 
the overlapping integrals between the wave functions localized on different
atoms, the better the method works.

In the case at hand, instead of a crystal field we have the  gauge 
field of the energon periodic in $\tau$.
The role of atoms is played by instantons and antiinstantons.
Instead of atomic wave functions localized on individual atoms
we have fermionic zero-modes localized on separate instantons and 
antiinstantons.
The non-relativistic Schr\"{o}dinger equation is replaced by
the Dirac equation. 
However, one can see that these distinctions do not play an essential role.
The decisive point is that in order to make the strong coupling method applicable
in our case, the fermionic wave function should be large only within 
well separated regions, where the gauge field is large.
We will see below that this is really the case.

The zero fermion mode, which exists in the field of the instanton (\ref{instanton}) localized at the origin,  reads
\begin{equation}
\psi^{(0)}_{\mathrm{in}}(x)=\frac 1 \pi \, \frac{ n_\mu}{(x^2+1)^{3/2}}
\left(
\begin{array}{c}
\sigma^+_\mu \epsilon \\ 0
\end{array} 
\right) ~.
\label{psi0-in}
\end{equation}
Here $n_\mu=x_\mu/\sqrt{x^2}$.  
The fermion wave function $\psi^{(0)}_{\mathrm{in}}(x)$ is written here as a chiral Dirac 4-spinor and isotopic
doublet. It possesses the index  $m=1,2$ (not written explicitly in (\ref{psi0-in})), which  distinguishes 
the two nonzero components in the Dirac spinor.
It also possesses the index $\alpha=1,2$ (also not shown in (\ref{psi0-in}) explicitly), which 
describes the isotopic doublet. These two indexes are blended together
in the matrix $\sigma^+_\mu \epsilon \equiv (\sigma^+_\mu \epsilon)_{m,\alpha}$, where $\epsilon$ is the $2\times 2$ antisymmetric matrix, $\epsilon_{12}=-\epsilon_{21}=1$.

We also need the fermion zero mode of the antiinstanton. Presuming temporally that the 
antiinstanton  is located at the origin we can write this mode as follows
\begin{equation}
\psi^{(0)}_{\mathrm{a-in}}=\frac 1 \pi \, \frac{ n_\mu}{(x^2+1)^{3/2}}
\left(
\begin{array}{c}
0 \\ \sigma^-_\mu \epsilon
\end{array} 
\right)~. 
\label{psi0-anti}
\end{equation}
Following the spirit of the strong coupling approach \cite{Ashcroft-Mermin}
we take a linear combination of 
the instanton zero modes localized at all those points,
where the energon potential is close to the potential created by the instanton
%(remember discussion of the instanton-antiinstanton chain)
\begin{equation}
\Psi_{\omega, \mathrm{in}}(\tau)= \frac 1 {\sqrt{N} }\, \sum_s e^{i\,s\, \omega\,T} \psi^{(0)}_{\mathrm{in}}(\tau-s\,T)
\label{omega-in}~.
\end{equation}
The coefficients $\exp (\,i s\, \omega T)$ in this linear combination are completely defined by Bloch's conditions
(\ref{Bloch}), $N$ is a large integer, which in the final stage of calculations will be taken to the limit
$N\rightarrow \infty$. Summation over an integer $s$ covers the region $-N/2<s<N/2$.
The parameter $\omega$ in the intermediate calculations  should satisfy $\omega=2\pi n/N$, where $n$ is an integer,
but in the limit $N,n\rightarrow \infty$ this parameter can take any value inside the Brillouin zone (\ref{Brillouin}).
For simplicity of presentation the argument ${\bf r}$ of the zero mode in (\ref{omega-in}) is suppressed.
One immediately verifies that $\Psi_{\omega, \mathrm{in}}(\tau)$ satisfies Bloch's periodic condition (\ref{Bloch}).

Similarly, we construct the fermion wave function using the antiinstanton zero modes
\begin{equation}
\Psi_{\omega, \mathrm{a-in}}(\tau)=\frac 1 {\sqrt{N} }\, \sum_s e^{i\,(s+\frac 12)\, \omega\,T} \psi^{(0)}_{\mathrm{a-in}}
\big( \tau-(s+\frac 12)\,T\big)~.
\label{omega-a-in}
\end{equation}
The argument $\tau -(s+\frac 12)T$ of the wave function of fermionic zero modes takes into account the fact
that the energon potential is close to the potential of the antiinstanton
in the vicinities of each point $\tau=(s+\frac 12)T$. This fact makes it also convenient to present
$s$ in the argument of the exponent function in (\ref{omega-a-in}) as $(s+\frac 12)\omega T$.
One verifies that the function $\Psi_{\omega, \mathrm{a-in}}(\tau)$ complies
with Bloch's condition. 

We have now to account for the discrete R-symmetry,
which is present in the energon potential.
The commutation of the Dirac operator 
with  the operator ${\hat R}$ stated in (\ref{nabla-invariant})
implies that the functions used to diagonalize the Dirac operator should also be eigenfunctions
of the operator $\hat R$.
Meanwhile the functions 
$\Psi_{\omega, \mathrm{in}}(\tau)$ and $\Psi_{\omega, \mathrm{a-in}}(\tau)$ 
are not the eigenfunctions of $\hat R$ because their transformations read
\begin{align}
&\hat R \,\Psi_{\omega, \mathrm{in}}\,=\,e^{i\,\omega\,T/2}\,\Psi_{\omega, \mathrm{a-in}}(x)~,
\label{ha1 R Psi}
\\
&\hat R \,\Psi_{\omega, \mathrm{a-in}}\,=\,e^{i\,\omega\,T/2}\,\Psi_{\omega, \mathrm{in}}(x)~.
\label{ha2 R Psi}
\end{align}
To remedy this deficiency take the linear combinations
\begin{equation}
\Psi_{\omega,\pm}\,=\,\frac 1 {\sqrt 2 }(\Psi_{\omega, \mathrm{in}}\pm \Psi_{\omega, \mathrm{a-in}})~.
\label{plus-minus}
\end{equation}
Using (\ref{ha1 R Psi})  and (\ref{ha2 R Psi}) one finds for these functions
\begin{equation}
\hat R\, \Psi_{\omega,\pm}\,=\,\pm e^{i\,\omega \,T/2} \Psi_{\omega,\pm}~.
\label{eigenvalue R}
\end{equation}
Deriving (\ref{eigenvalue R})  one takes into account that the operator of inversion
changes the chirality of 4-spinors  as illustrated by the identity
\footnote{Equation (\ref{P-spinor}) specifies our choice of the phase for the operator of inversion, 
compare Chapter 19 of Ref. \cite{LL4}.}
\begin{equation}
\hat P
\left(
\begin{array}{c}
1 \\ 0
\end{array} 
\right)
\,=\,
\left(
\begin{array}{c}
0 \\ 1
\end{array} 
\right).
\label{P-spinor}
\end{equation}
Equation (\ref{eigenvalue R}) states that $\Psi_{\omega,\pm}$ 
are eigenfunctions of $\hat R$, 
precisely as required by the symmetry condition (\ref{omega,pm}).
The discussion presented after (\ref{omega,pm}) implies that
Eq.(\ref{eigenvalue R}) automatically ensures the validity of Bloch's
conditions (\ref{Bloch}).

One can see that there is only one possible way for constructing 
a fermionic wave function, which is 
a linear combination of the fermion zero modes related to instantons and antiinstantons,
and which satisfies the necessary symmetry conditions.
The function in question $\Psi_{\omega,\sigma}$ from (\ref{plus-minus})
possesses two parameters, $\omega$ and $\sigma=\pm 1$.
In other words, for a chosen parameters $\omega,\sigma$
the function is unique. 
As a result the strong coupling method
greatly simplifies.
The expression for the eigenvalue in this case simply reads 
\footnote{Compare Eq.(\ref{H/N}) of this work with Eq.(10.15) of \cite{Ashcroft-Mermin}. 
%which is valid when
%only one atomic wave function is taken into account. 
The notation is quite different, but careful comparison shows their close resemblance.}
\begin{equation}
\varepsilon_{\omega,\pm}\,=\,\frac{\langle\, \Psi_{\omega,\pm}\,|i\,\gamma_\mu \,\nabla_\mu|\,\Psi_{\omega,\pm}\,\rangle}
{\langle \, \Psi_{\omega,\pm}\,|\,\Psi_{\omega,\pm}\,\rangle}\,\equiv\,\frac{\mathcal{H}}{\mathcal{N}}~.
\label{H/N}
\end{equation}
Here $\mathcal{H}$ and $\mathcal{N}$ equal the nominator and denominator of the expression in the middle.
Notation in (\ref{H/N}) presumes that matrix elements are calculated as integrals over the 4D Euclidean space.

Straightforward calculations, see (\ref{N-norma}) in Appendix \ref{Normalization}, show that
\begin{equation}
{\mathcal{N}}\,=\,{\langle \, \Psi_{\omega,\pm}\,|\,\Psi_{\omega,\pm}\,\rangle}\,=\,1+O(\beta^2)~.
\label{NN}
\end{equation}
Hence, hunting for the main term in an expansion of $\varepsilon_{\omega,\pm}$ over $\beta$, 
we can restrict ourselves to a trivial approximation ${\mathcal{N}}\approx 1$. 
This complies with conditions of applicability of the strong coupling approximation,
which requires the overlapping integrals between functions located at different 
centers be small.

Thus, in order to find the eigenvalues $\varepsilon_{\omega,\pm}$ we need only to calculate 
the matrix element $\mathcal{H}$.
First of all, using the fact that $\gamma_\mu$ matrices mix chirality 
we can rewrite it as follows
\begin{equation}
{\mathcal{H}} \,=\,
\pm\,
\mathrm{Re} \,\langle \,\Psi_{\omega,\mathrm{in}}\,|\,i\,\gamma_\mu \,\nabla_\mu|\,\Psi_{\omega,\mathrm{a-in}}\,\rangle~,
\label{nominator}
\end{equation}
where the sign in front complies with the sign in the chosen function $\Psi_{\omega,\pm}$.
Secondly, using Eqs.(\ref{omega-in}) and (\ref{omega-a-in}), we rewrite this expression once again 
\begin{equation}
\langle \,\Psi_{\omega,\mathrm{in}}\,|\,i\,\gamma_\mu \,\nabla_\mu|\,\Psi_{\omega,\mathrm{a-in}}\,\rangle=\sum_s e^{i\big(s+\frac 12\big)\, \omega T} h_s~,
\label{h-s}
\end{equation}
where
\begin{equation}
h_s\,=\langle \,\psi^{(0)}_{\mathrm{in}}(\tau)|\,i\gamma_\mu \,\nabla_\mu|\,\psi^{(0)}_{\mathrm{a-in}}\big(\tau-(s\!+\!\frac 12\,)\,T\,\big)\,\rangle~.
\label{h-s-again}
\end{equation}
In Eq.(\ref{h-s}) we take the desired limit $N\rightarrow \infty$ and presume accordingly 
that $-\infty< s<\infty$.

The main contribution to the integral 
over $\tau$ in (\ref{h-s-again})  comes from the region located between the origin, $\tau=0$, 
where the function $\psi^{(0)}_{\mathrm{in}}({\bf r},\tau)$
is large and the point 
$\tau=(s+\frac 12)T$, where the function
$\psi^{(0)}_{\mathrm{a-in}}({\bf r},\tau-(s+\frac 12)T)$ is prominent.
Moreover, it is shown in Appendix \ref{Matrix element}
that the integral over $\tau$ collapses to the intermediate point 
$\tau=\tau_0=\frac12(s+\frac12)T$. This is stated in Eq.(\ref{d3rt0}),
which we reiterate here
\begin{equation}
h_s=i\,\mathrm{sign}\big(s+\frac 12\big)\int 
\psi^{(0)+}_{\mathrm{in}}(\tau_0)\,\gamma_4 \,\psi^{(0)}_{\mathrm{a-in}}(-\tau_0)\,d^3r.
\label{h-s-g4}
\end{equation}
Observe that both functions
$\psi^{(0)}_{\mathrm{in}}$ and $\psi^{(0)}_{\mathrm{a-in}}$ appear
in (\ref{h-s-g4}) only for large values of the variable $\tau$, $\tau=\pm\tau_0$, where the potential created by the
energon is small because $|\tau_0|\gg \rho$. Correspondingly, in this region the asymptotic 
forms of the equations for the fermion wave function are satisfied by
both the instanton and antiinstanton modes, 
$\gamma_\mu \partial_\mu \,\psi^{(0)}_{\mathrm{in}}\approx 
\gamma_\mu \partial_\mu \,\psi^{(0)}_{\mathrm{a-in}} \approx 0$.
This shows that the wave functions in (\ref{h-s-g4}) are used only within the region where
their validity is justified.
Equation (\ref{h-s-g4}) can be compared with the known Holstein-Herring method, 
also called Smirnov's method, which is commonly used in molecular 
and condense matter calculations \cite{Holstein,Herring,Smirnov}, 
for a brief description see also chapters 50 and 81 of \cite{Landau}.

Direct calculations in (\ref{h-s-g4}), see (\ref{h-s-finally}) in Appendix \ref{Matrix element}, give
\begin{equation}
h_s\,\approx\,-\,\frac{4\,i\,\beta^3}{(s+\frac 12)^3}~.
\label{h-s-result}
\end{equation}
One substitutes now (\ref{h-s-result}) into (\ref{h-s}) and rearranges the summation 
over $s$ there to make it running over $s\ge 0$. 
After that Eqs.(\ref{H/N}), (\ref{NN}) and (\ref{nominator}) give 
\begin{equation}
\varepsilon_{\omega,\pm}\,\approx\,\pm \,8 \,\beta^3\,\sum_{s=0}^\infty \,\frac{ \sin \Big(\,\big( s+\frac 12)\,\omega\,T\Big)}
{ (s+\frac 1 2 )^3}~.\label{E=8etc}
\end{equation}
Using equation 5.4.6.13 from Ref. \cite{Prudnikov} we finally derive a simple, appealing result
\begin{align}
&\varepsilon_{\omega,\pm}\,=\,\pm \,4\,\pi^2\beta^3\,\omega T\,\Big(1-\frac{|\omega| T}{2\pi}\Big)~.
\label{answer}
\end{align}
Here $\omega\,T$ belongs to the first Brillouin  zone (\ref{Brillouin}) and 
$\beta=\rho/T$ is the small gas parameter.
Equation (\ref{answer}) is one of the important results of this work.
Its discussion is given in Section \ref{Discussion}.

\section{Discussion}
\label{Discussion}
Consider properties of the eigenvalues $\varepsilon_{\omega,\pm}$ of the Dirac operator 
presented in Eq.(\ref{answer}).
Our initial idea was that conditions  $\mathcal{E}<0$, $|\mathcal{E}|\rho \ll 1$
should bring into existence fermion modes
with properties similar to fermion zero modes of 
instantons and antiinstantons. 
Equation (\ref{answer}) complies with these expectations.
The eigenvalues found there are  small, $\propto \beta^3$
\footnote{This result agrees with a preliminary estimate obtained previosly with J.Diacoumis.}.
The smallness of the eigenvalues also justifies the validity of the strong coupling approach, which
was implemented in Eqs.(\ref{plus-minus}).
%An alternative way to verify it  validity provides the normalization conditions.
%According to Eq.(\ref{NN}) the deviation of the normalization
%factor $\mathcal{N}$ from unity is small, which complies with
%the strong coupling approach, see \cite{Ashcroft-Mermin}). 
%This property is related to the fact
%that the wave functions in (\ref{omega-in}), (\ref{omega-a-in}) are
%large only in those regions of the 4D Euclidean space, where the gauge field is strong.
%This anticipated from the very beginning property makes the 
%strong coupling approach the more accurate, the smaller $\beta$
%in (\ref{beta}) is.
The eigenfunctions (\ref{plus-minus}) are localized in 3D coordinate
space, $\Psi_{\omega,\pm}({\bf r},\tau)\rightarrow 0$ when ${\bf r}\rightarrow \infty$, 
in agreement with Eq.(\ref{r infty}). This property is similar to the behavior of 
the instanton zero modes.

However, there is a notable
distinction between $\Psi_{\omega,\pm}({\bf r},\tau)$ and conventional
fermionic zero modes, which appear for self-dual gauge fields.
The latter decrease for large $\tau$, while the former satisfy 
Bloch's periodic condition (\ref{Bloch}), which 
originates from the periodic nature of the
gauge field created by the energon. 
%and prompts an appearance of the zone structure,
%which manifests itself as an $\omega$-dependence of the eigenvalue problem.

It is well known, see e.g. \cite{Ashcroft-Mermin},
that on the boundaries of the Brillouin zone the derivative of the
eigenvalue should vanish. In our case this condition reads
\begin{equation}
\frac{\partial\varepsilon_{\omega,\pm}}{\partial \omega} \,=\,0, \quad \frac{\omega T} {\pi}=\pm 1~.
\label{B-boundary}
\end{equation}
It is gratifying  that Eq.(\ref{answer}) complies with this necessity.

Notably, the spectrum of eigenvalues of the Dirac operator
possesses the doubly degenerate zero eigenvalue, $\varepsilon_{0,\pm}=0$, 
which is present at the center of the Brillouin zone $\omega=0$. 
Equation (\ref{eigenvalue R}) for $\omega=0$ shows that
the two corresponding fermion states are transformed
under the $\hat R$ operator as follows
\begin{equation}
\hat R\,\Psi_{0,\pm}\,=\,\pm \Psi_{0,\pm}~,
\label{R-pm}
\end{equation}
It is tempting to conjecture that
the very presence of states with zero eigenvalues of the Dirac operator
in the fermion spectrum
can be attributed to the  existence of the R-symmetry (\ref{R-invariant})
for the energon solution.
It would be interesting to justify this point on the basis of arguments of a general nature, 
appealing probably to the Atiya-Singer theorem.
However, we will not make such attempt in the present work
since we restrict our techniques here to the simple strong coupling approach.
(The importance of the R-symmetry can be probed in this technique
by applying the method of strong coupling to a set of functions, which 
do not satisfy the symmetry conditions (\ref{omega,pm}), for example taking 
a pair of functions $\Psi_{\omega, \mathrm{in}},\Psi_{\omega, \mathrm{a-in}}$
from (\ref{omega-in}),(\ref{omega-a-in}). One verifies that in this case
the spectrum of eigenvalues of the Dirac operator has no zero eigenvalues.
This simple argument can be considered as a hint, 
more accurate consideration is needed.)
%At last, one should ask oneself, if the found spectrum of fermions
%can be used elsewhere, in theory, or even in experimental applications.

The results reported warrant further investigation.
It would be  interesting to consider properties of the fermion spectrum
in the energon field in general terms, 
possibly proving the presence of zero eigenvalues
of the Dirac operator for an arbitrary energon configuration.
The periodic in $\tau$ nature of the energon field
makes such studies an appealing task by itself, whereas 
an interplay of ideas with solid state physics
%, Bloch periodic conditions, Brillouin zone,
provides additional impetus.
The existence of zero eigenvalues of the Dirac operator inspires
consideration of  creation of fermion-antifermion pairs 
in the field of the energon, similar to pair
creation by instantons. 

%Summarizing, the spectrum of states, which fermions possess in the gauge field 
%created by the energon is evaluated for a particular situation, 
%when parameters of the energon are tuned in such a way that the gauge 
%field is close to the field created 
%by a periodic chain of conventional instantons and anti-instantons.
%The spectrum exhibits several interesting features, 
%incorporating in particular the states for which 
%the Dirac operator possesses the zero eigenvalue.

\appendix
\section{Normalization}
\label{Normalization}
Consider the normalization factor $\mathcal{N}$, which arises in the denominator of Eq.(\ref{H/N}). Using Eq.(\ref{plus-minus})
and then Eqs.(\ref{omega-in}), (\ref{omega-a-in}) together with (\ref{psi0-in}) and (\ref{psi0-anti}) we present it as follows
\begin{align}
&{\mathcal{N}}\,=\,{\langle \, \Psi_{\omega,\pm}\,|\,\Psi_{\omega,\pm}\,\rangle}\,=\,
{\langle \, \Psi_{\omega,\mathrm{in}}\,|\,\Psi_{\omega,\mathrm{in}}\,\rangle}\,=\,
\nonumber
\\
&\sum_{s=-\infty}^{\infty} e^{i\,s\,\omega\,T}
{\langle \, \psi^{(0)}_{\mathrm{in}}(\tau)\,|\,\psi^{(0)}_{\mathrm{in}}(\tau-sT)\,\rangle}\,=\,
\nonumber
\\
&\sum_{s=-\infty}^{\infty} e^{i\,s\,\omega\,T}\,
\frac{2}{\pi^2} \int \frac{n_\mu n'_\mu}{(x^2+1)^{3/2}(x'^2+1)^{3/2}}\,d^4x
\label{N=1}
\end{align}
Here $x=({\bf r},\tau)$, $n_\mu=x_\mu/\sqrt{x^2}$, and $x'_\mu=({\bf r}, \tau-s T)$, $n'_\mu=x'_\mu/\sqrt{x'^2}$.
Straightforward calculation of the integral in (\ref{N=1}) and obvious rearrangements in the summation
(to make it running over positive $s$ only) gives
\begin{align}
&{\mathcal{N}}\,\approx\,1+4\beta^2\sum_{s=1}^\infty\,\frac{\cos (s\,\omega T)}{s^2}\,=\,
\nonumber
\\
&1+4\pi^2\,\beta^2\,\Big( \Big( \frac{\omega T}{2\pi} \Big)^2-\frac{|\omega| T}{2\pi}+\frac16\Big)\,.
\label{N-norma}
\end{align}
The last identity here is written using equation 5.4.2.7 from 
\cite{Prudnikov}. 
%Since $\beta$ is small, the most important role in Eq.(\ref{N-norma}) belongs to the unity term,
%which complies with the spirit of strong coupling approach \cite{Ashcroft-Mermin}.

\section{Matrix element}
\label{Matrix element}
Let us simplify the matrix element $\mathcal{H}$, 
which appears in the nominator of (\ref{H/N}),
by calculating $h_s$ from Eqs.(\ref{h-s}).
Presuming firstly that $s\ge 0$ 
let us rewrite (\ref{h-s-again}) as follows
\begin{align}
&h_s=\int d^3r\Bigg(\int_{-\infty}^{\tau_0}d\tau+\int_{\tau_0}^{\infty}d\tau\Bigg)\times
\label{t0}
\\
&\Big(\psi^{(0)}_{\mathrm{in}}({\bf r},\tau)\Big)^+ \,i\gamma_\mu \nabla_\mu\ 
\psi^{(0)}_{\mathrm{a-in}}\Big({\bf r},\tau-\Big(s+\frac12\Big)T\Big)~.
\nonumber
\end{align}
Here an obvious bridging notation is used for integrals over $\tau$.
It is convenient to choose the intermediate integration limit $\tau_0$ 
strictly between the origin and the point $(s+\frac12)T$,
\begin{equation}
\tau_0=\frac12\Big(s+\frac12\Big)\,T~.
\label{tau0}
\end{equation}
The idea behind Eqs.(\ref{t0}), (\ref{tau0}) is that the integral over $\tau$
is saturated in the vicinity of $\tau_0$. 
Similar matrix elements have long being used in molecular and solid state physics \cite{Holstein,Herring,Smirnov}.
For the region $-T/4<\tau<\tau_0$ the potential created by the energon
is either small, or (if large) is
close to the potential created by the instanton located at the origin.
The first option takes place  when $r^2+\tau^2>1$, the second when
$r^2+\tau^2\le 1$. In both cases we can approximate the potential
created by the energon by the instanton potential and consequently write
$\nabla_\mu \approx \nabla_{\mu,\mathrm{in}}$.
Meanwhile, the region $\tau<-T/4$ is of no importance since both
wave functions from (\ref{t0}) are small there.
Thus, an approximation $\nabla_\mu \approx \nabla_{\mu,\mathrm{in}}$ is 
applicable for all $\tau<\tau_0$ .
Similarly, in the region $\tau_0<\tau$ we can approximate 
the energon potential by the potential created by the antiinstanton, which
is located at ${\bf r}_\mathrm{a-in}=0$ and $\tau_\mathrm{a-in}=T/2$.
In this region one writes therefore
$\nabla_\mu \approx \nabla_{\mu,\mathrm{a-in}}$. 
This last relation implies that the second integral over $\tau$ in (\ref{t0}),
the one from $\tau_0$ to $\infty$, turns zero
because the relevant  Dirac operator annihilates the antiinstanton
zero mode $\gamma_\mu\nabla_{\mu,\mathrm{a-in}} \psi^{(0)}_{\mathrm{a-in}}=0$.

Hence we need to consider in  (\ref{t0}) only the first integral over $\tau$,
which runs over the interval  $-\infty<\tau <\tau_0$.
Integrating over $\tau$ by parts and using the fact that the relevant Dirac operator annihilates the
fermion zero mode,
$\gamma_\mu\nabla_{\mu,\mathrm{in}} \psi^{(0)}_{\mathrm{in}}=0$, we obtain
\begin{equation}
h_s=
i~\mathrm{sign}\Big(s+\frac12\Big)\int
\Big(\psi^{(0)}_{\mathrm{in}}({\bf r},\tau_0)\Big)^+ \!
\gamma_4
\psi^{(0)}_{\mathrm{a-in}}({\bf r},-\tau_0)d^3r.
\label{d3rt0}
\end{equation}
%So far we considered the case $s\ge 0$. 
For $s<0$ the result looks very similar, but the sign in front of the final expression changes.
Having this in mind, the factor $\mathrm{sign}(s+\frac12)$ was incorporated in Eq.(\ref{d3rt0}) 
to make it applicable for all $s=0,\pm1,\dots\, $.
We see that in accord with our preliminary expectations
the integral over $\tau$ collapses to the point $\tau=\tau_0$.

Further calculations are straightforward. Using Eqs. (\ref{psi0-in}) and (\ref{psi0-anti}) 
one obtains
\begin{align}
&\int\Big(\psi^{(0)}_{\mathrm{in}}({\bf r},\tau_0)\Big)^+ 
\gamma_4\,
\psi^{(0)}_{\mathrm{a-in}}({\bf r},-\tau_0)\,d^3r=
%\nonumber
\label{Tr}
\\
&\frac{1}{\pi^2}\int \,
\mathrm{Tr}(\sigma_\mu^-\sigma_\nu^-)\,
\frac{n_\mu n'_\nu}{(x^2+1)^{3/2}(x'^2+1)^{3/2} }\,d^3r=
\nonumber
\\
&
-\frac{2}{\pi^2}\int \,\frac{d^3r}{(x^2+1)^3}=-\frac 1{2(\tau_0^2+1)^{3/2}}\approx
\nonumber
%\\ &
-\frac{ 4\beta^3 \mathrm{sign}(s+\frac12)} { (s+\frac 1 2 )^3 } ~.
\end{align}
Here the intermediate expressions incorporate the variables 
$x_\mu=({\bf r},\tau_0)$,  $n_\mu=x_\mu/\sqrt{x^2}$,
as well as $x'_\mu=({\bf r},-\tau_0)$ and $n'_\mu=x'_\mu/\sqrt{x'^2}$.
We find from these conditions that
\begin{equation}
\sigma_\nu^-  n'_\nu\,=\,-\sigma_\nu^+  n_\nu~,
\label{sigma-+}
\end{equation}
and consequently
\begin{equation}
\mathrm{Tr}\, (\sigma_\mu^-\sigma_\nu^-)\,n_\mu n'_\nu\,=\,
-\mathrm{Tr}\, (\sigma_\mu^-\sigma_\nu^+)\,n_\mu n_\nu\,=\,-2~.
\label{trace}
\end{equation}
In the third line in (\ref{Tr}) we use the fact that
$x^2=x'^2$. The last expression in (\ref{Tr}) is written remembering
(\ref{tau0}) and condition $\beta=1/T\ll 1$.
Finally, from (\ref{d3rt0}) and (\ref{Tr}) we derive
\begin{equation}
h_s\,\approx\,-\,\frac{4\,i\,\beta^3}{(s+\frac 12)^3}~,
\quad \quad s=0,\pm 1,\dots .
\label{h-s-finally}
\end{equation}

\section*{Acknowledgment} 
Discussions with James Diacoumis are appreciated. The work was supported by the Australian Research Council. 

%\bibliography{ENER}
%%\bibliographystyle{nature}
%\bibliographystyle{h-physrev}
%%\bibliographystyle{Nunsrt}
%%\bibliographystyle{hunsrt}
%%\bibliographystyle{h-physrev}
%%plain
%%nature
%%is-unsrt
%%unsrt
%%Nunsrt
%%nar
%%phcpc
%%phiaea
%%hunsrt
%%h-physrev
%\end{document}

\end{document}